\documentclass[sigconf, nonacm]{acmart}

\AtBeginDocument{%
  }

\setcopyright{none} 
\copyrightyear{2024}
\acmYear{2024}
\acmDOI{XXXXXXX.XXXXXXX}

\acmConference['ISCA 2025']{The 52nd IEEE/ACM International Symposium on Computer Architecture}{June 21--25, 2025}{Tokyo, Japan}
\acmISBN{978-1-4503-XXXX-X/18/06}

\usepackage[most]{tcolorbox}

\newtcbtheorem{Summary}{\bfseries Summary}{enhanced,drop shadow={black!50!white},
  coltitle=black,
  top=0.3in,
  attach boxed title to top left=
  {xshift=1.5em,yshift=-\tcboxedtitleheight/2},
  boxed title style={size=small,colback=lightgray}
}{summary}

\newtcolorbox[auto counter]{summary}[1][]{title={\bfseries Summary~\thetcbcounter},enhanced,drop shadow={black!50!white},
  coltitle=black,
  top=0.1in,
  left=-0.1in,
  attach boxed title to top left=
  {xshift=0.1em,yshift=-\tcboxedtitleheight/2},
  boxed title style={size=small,colback=pink},#1}

\usepackage{float}

\usepackage{stfloats}

\newcommand\bra[1]{\left\langle#1\right|}
\newcommand\ket[1]{\left|#1\right\rangle}

\usepackage{amsmath,mathtools, amsthm}

\newcommand\todo[1]{\textcolor{black}{#1}}
\newcommand\todoo[1]{\textcolor{black}{#1}}

\usepackage{xcolor}
\usepackage{color}
\usepackage{listings}
\usepackage{multirow}
\usepackage{multicol}
\usepackage{url}
\usepackage{tikz}

\begin{document}

\title{Optimizing Quantum Communication for Quantum Data Centers with Reconfigurable Networks}
\author{Hezi Zhang}
\email{hezi@ucsd.edu}
\affiliation{
 \institution{University of California}
 \city{San Diego}
 \country{USA}
}
\author{Yiran Xu}
\email{yix072@ucsd.edu}
\affiliation{
 \institution{University of California}
 \city{San Diego}
 \country{USA}
}
\author{Haotian Hu}
\email{hah041@ucsd.edu}
\affiliation{
 \institution{University of California}
 \city{San Diego}
 \country{USA}
}
\author{Keyi Yin}
\email{keyin@ucsd.edu}
\affiliation{
 \institution{University of California}
 \city{San Diego}
 \country{USA}
}
\author{Hassan Shapourian}
\email{hshapour@cisco.com}
\affiliation{
 \institution{Cisco Quantum Lab}
 \city{San Jose}
 \country{USA}
}
\author{Jiapeng Zhao}
\email{penzhao2@cisco.com}
\affiliation{
 \institution{Cisco Quantum Lab}
 \city{San Jose}
 \country{USA}
}
\author{Ramana Rao Kompella}
\email{rkompell@cisco.com}
\affiliation{
 \institution{Cisco Quantum Lab}
 \city{San Jose}
 \country{USA}
}
\author{Reza Nejabati}
\email{rnejabat@cisco.com}
\affiliation{
 \institution{Cisco Quantum Lab}
 \city{San Jose}
 \country{USA}
}
\author{Yufei Ding}
\email{yufeiding@ucsd.edu}
\affiliation{
 \institution{University of California}
 \city{San Diego}
 \country{USA}
}
 


\begin{abstract}

Distributed Quantum Computing (DQC) enables scalability by interconnecting multiple QPUs. Among various DQC implementations, quantum data centers (QDCs), which utilize reconfigurable optical switch networks to link QPUs across different racks, are becoming feasible in the near term. However, the latency of cross-rack communications and dynamic reconfigurations poses unique challenges to quantum communication, significantly increasing the overall latency and exacerbating qubit decoherence.
In this paper, we introduce a new optimization space to parallelize cross-rack communications and avoid frequent reconfigurations, which incurs additional in-rack communications that can be further minimized. Based on this, we propose a flexible scheduler that improves communication efficiency while preventing deadlocks and congestion caused by the flexibility.
Through a comprehensive evaluation, we show that our approach reduces the overall latency by a factor of $\todo{8.02}$, thereby mitigating qubit decoherence, with a small overhead.

\end{abstract}





\maketitle

\section{Introduction}
\label{sect:introduction}

Distributed quantum computing (DQC)\cite{cacciapuoti2019quantum, AndresMartinez2019AutomatedDO, cuomo2020towards, laracuente2022modeling} has emerged as a promising approach to address scalability challenges in quantum computing. By enabling quantum communication between multiple quantum processing units (QPUs), DQC extends the capabilities of single-chip systems to multi-node architectures. This approach allows scalability in two complementary ways: scaling up individual nodes by enhancing the capabilities of each QPU and scaling out by connecting multiple nodes to work collaboratively. Achieving this dual scalability unlocks the potential for significantly larger and more powerful quantum computing systems.

\begin{figure}[h!]
        \centering
        \includegraphics[width=0.95\linewidth]{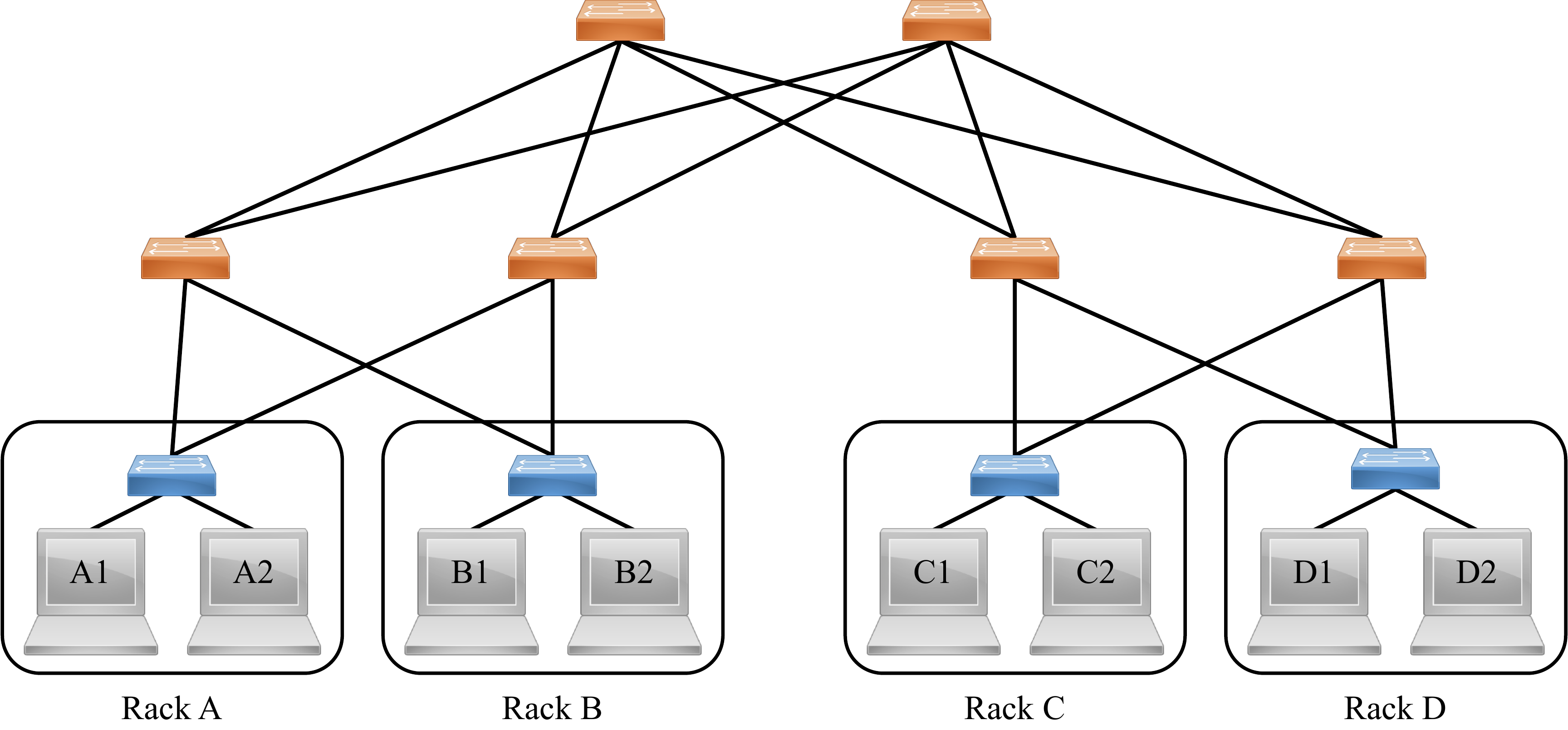}
        
        \caption{A QDC with QPUs connected by a local optical network.}
        \label{fig:intro}
\end{figure}

Among various implementations of DQC, quantum data centers (QDCs) \cite{cuomo2020towards, awschalom2021development, caleffi2022distributed, ang2022architectures} have gained significant attention from industry as a practical near-term solution. Unlike early visions of DQC that focused on long-distance quantum communication across cities, QDCs place multiple racks of QPUs within a single room or facility, interconnected through a local optical network. Their controlled environment and relatively short communication distances mitigate photon loss and environmental noise, aligning them better with near-term computational needs.
The optical network of QDCs, comprising optical fibers and reconfigurable optical switches~\cite{nejabati2022dynamic, bahrani2023analysing}, dynamically adjusts switch ports during computation to establish direct optical channels between QPUs. As shown in Fig.~\ref{fig:intro}, most inter-rack switches (core switches), depicted in orange, can leverage existing classical optical switch technology, while only a small number of specialized quantum switches, depicted in blue, need to be deployed on top of each rack (ToR switches) to enable EPR pair generation for quantum communication. This significantly enhances the practicality and cost-efficiency of QDCs, making them a more viable infrastructure for DQC deployment in the near future.

\begin{figure}[h!]
        \centering
        \includegraphics[width=0.33\linewidth]{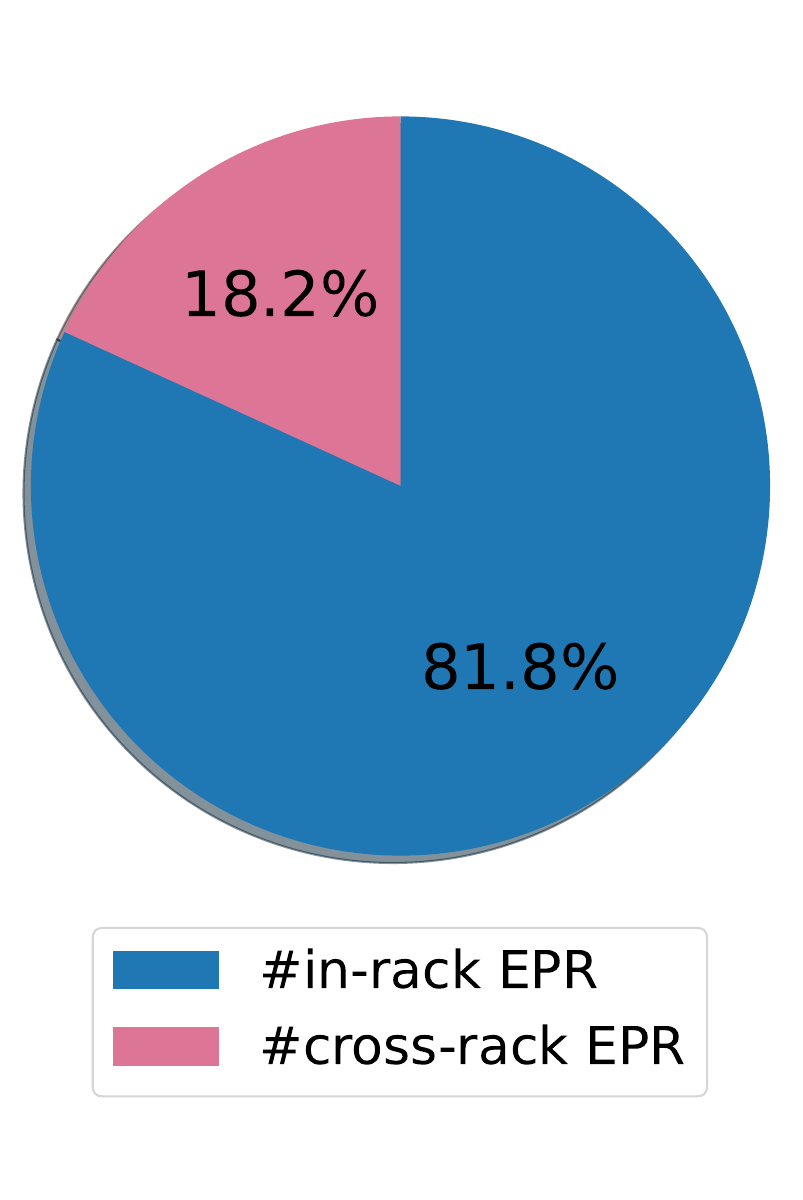}
        \hspace{8pt}\includegraphics[width=0.33\linewidth]{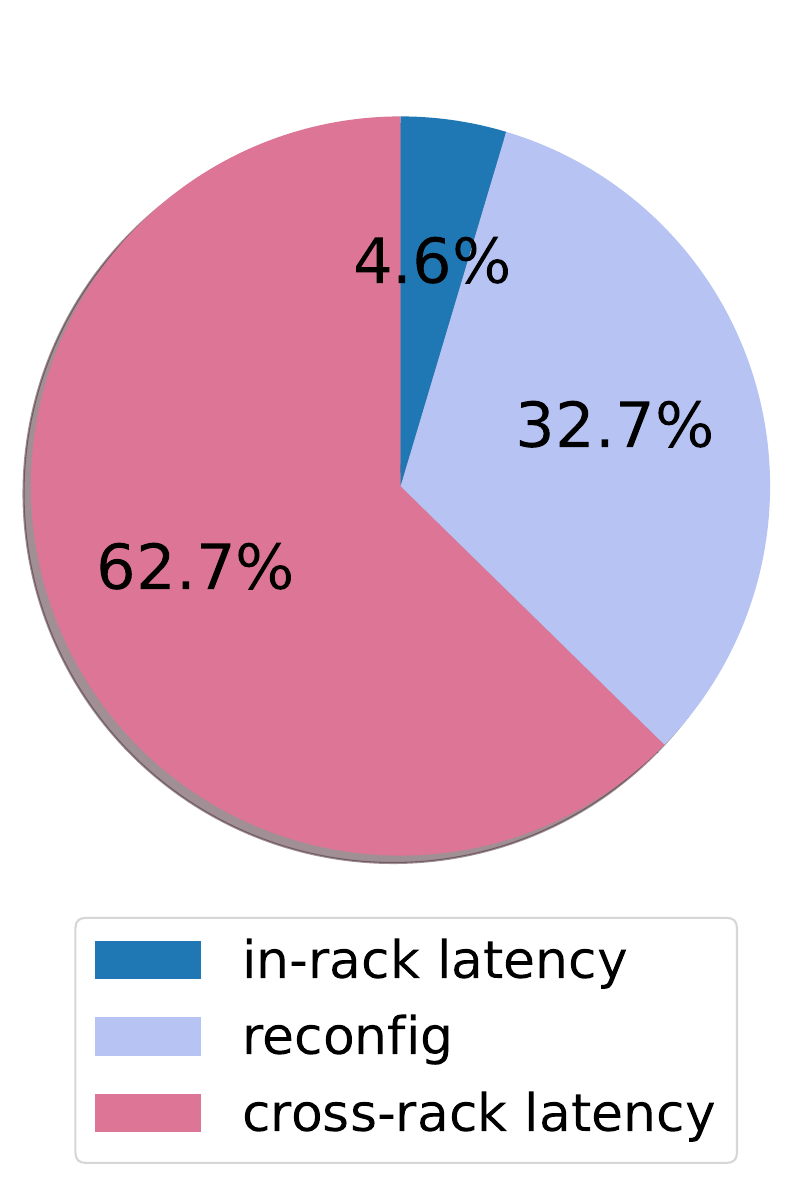}\\
        (a)\hspace{85pt}(b)
        \vspace{-5pt}

        \caption{Number percentages of in-rack and cross-rack EPR pairs (a) and latency percentages of in-rack EPR pair generation, reconfiguration and cross-rack EPR pair generation.}
        \label{fig:profiling}
\end{figure}

The primary bottleneck in DQC lies in inter-QPU communication, which is significantly slower and more error-prone than QPU computation. Existing software solutions~\cite{
diadamo2021distributed,ferrari2021compiler,haner2021distributed,AutoComm,QuComm} aim to reduce latency and improve fidelity by minimizing the number of required EPR pairs used for communication. However, these approaches show substantial inefficiencies when applied to the QDC architecture due to two key factors.
\textbf{First}, the heterogeneity between classical and quantum switches require converters to facilitate their interaction. This added complexity, combined with relatively long distances between racks, results in higher latency ($\sim$10 ms) for inter-rack EPR pair generation than in-rack EPR pair generation ($\sim$0.1 ms).  \textbf{Second}, each switch reconfiguration introduces about 1 ms of additional latency, which is necessary to minimize photon loss. As illustrated in Fig.~\ref{fig:profiling}, profiling results of existing software show that while inter-rack EPR pairs constitute only 18.2\% of the total required EPR pairs, they account for 62.7\% of the overall latency, with reconfigurations contributing an additional 32.7\%. This implies that inter-rack communication and frequent reconfigurations substantially increase latency, which in turn, also degrades fidelity due to the increased qubit decoherence over time.

To address the unique challenges in QDCs, innovative compilation techniques are required. Our key insight is that the heterogeneous architecture of QDCs actually offers a significant optimization space for hiding communication latencies across different QPUs. Traditionally, a QPU's communication latency could only be hidden among its own computation time, as all inter-QPU communications are almost equally costly. However, the heterogeneity in QDCs allows for a more strategic approach: we can hide the latencies of more expensive cross-rack communications by overlapping them,
at the cost of incurring additional, less expensive in-rack communications. By leveraging this optimization strategy, we can reduce overall latency and enhance fidelity of program execution, as long as the overhead from additional in-rack communications can be kept to a minimum.

At a high level, this latency hiding is achievable for two key reasons.
First, EPR pairs can be generated by merging multiple pre-existing EPR pairs via entanglement swapping. Unlike that in repeater networks \cite{briegel1998quantum,munro2015inside,azuma2022quantum}, entanglement swapping on QPUs can be performed fast and deterministically by gates and measurements. This allows the generation of a congested EPR pair to be split into several non-congested pairs, thereby fully utilizing the network bandwidth to enhance communication efficiency.
Second, in quantum systems, the preparation of EPR pairs, which dominates the latency of inter-QPU communication, can be decoupled from actual communications. This is feasible because EPR pairs do not carry any program data and can be buffered temporarily after their preparation \cite{QuComm}. This decoupling provides greater flexibility in the timing of EPR pair generations, allowing for the effective hiding of their latencies.

However, leveraging these features for compilation is highly non-trivial. On one hand, they can introduce fidelity overheads in two ways. First, splitting the preparation of EPR pairs necessitates generation of additional EPR pairs, which can compromise overall fidelity, as even in-rack EPR pairs are more error-prone than gates on QPUs. Second, decoupling the preparation of EPR pairs from communications requires storing these pairs in buffers, which brings a risk of decoherence. On the other hand, the flexibility enabled by these features can lead to complications such as deadlocks and buffer congestion due to hardware and bandwidth constraints. This calls for a careful design of scheduling principles and a retry mechanism whenever needed.

To this end, we present an efficient compiler that strikes a balance between latency reduction and fidelity overhead, while handling the problems of deadlock and buffer congestion. \textbf{First}, it parallelizes cross-rack EPR pair generations by dynamically changing their sources and destinations through EPR split, with the split resulting in a new cross-rack pair and some additional in-rack pairs (referred to as post-split EPR pairs). In this way, even if a cross-rack EPR pair can not be generated between the source and destination QPUs because one (or both) of them is busy, a substitute cross-rack EPR pair can be generated between the source and destination racks through another QPU in the same rack as the busy one. Once the busy QPU becomes available, it can communicate with the other QPU by generating an additional in-rack EPR pair promptly, thereby completing the generation of the original cross-rack EPR pair through an entanglement swapping.
\textbf{To mitigate the fidelity overhead} brought by additional in-rack EPR pairs, we prepare multiple copies of each post-split in-rack EPR pair and enhance its fidelity by incorporating an entanglement distillation \cite{bennett1996purification} using the extra copies. 

\textbf{Second}, we further hide the latency of generating the additional in-rack EPR pairs, either from split or for distillation, by a collective generation of them, so that they are prevented from becoming a new communication bottleneck. That is, given that the latency of reconfiguration is much longer than that of in-rack EPR pair generation, we can collect in-rack EPR pairs between the same QPUs and generate them in one shot when the channel between the QPUs is available. In this way, we reduce the average latency of in-rack EPR pair generation by avoiding frequent reconfiguration. The collection of in-rack EPR pair pairs is guided by a scheduler that looks ahead into program demand in the near future. This program-aware guidance reduces the wait time of both in-rack and cross-rack EPR pairs, thereby \textbf{minimizing the fidelity overhead} brought by storage of prepared EPR pairs.

\textbf{Third}, we propose a flexible scheduler that realizes the optimizations above while preventing potential deadlocks and buffer congestion caused by the flexibility. Specifically, we set several rules for split and scheduling of EPR pairs. These rules mitigate the risk of deadlock and buffer congestion by ensuring enough buffer size for future EPR pairs, despite that the problems can not be completely avoided.  In case that deadlock or buffer congestion does happen, we resolve it by incorporating an efficient retry mechanism that downgrades to more conservative scheduling  strategies when problems are detected.

\begin{itemize}
    \item We propose a compiler to overcome the communication challenges of QDC architecture while striking a balance with its fidelity overhead and avoiding problems of deadlock and buffer congestion.

    \item We improve the communication efficiency by hiding latencies of cross-rack communications into each other, at the cost of incurring additional in-rack communications, with the latency of in-rack communications further reduced by a collective generation.
    \item We minimize the fidelity overhead by incorporating entanglement distillation for the additional in-rack communications and reducing the wait time of prepared EPR pairs through a program-aware guidance.
    \item With a comprehensive evaluation based on simulations with practical hardware parameters, we demonstrate the significant outperformance over previous approaches in reducing communication latency while maintaining a low fidelity overhead.

\end{itemize}

\section{Background and Related Work}\label{sect:background}

\subsection{Distributed Quantum Computing (DQC)}
\begin{figure}[h!]
        \centering
        \includegraphics[width=\linewidth]{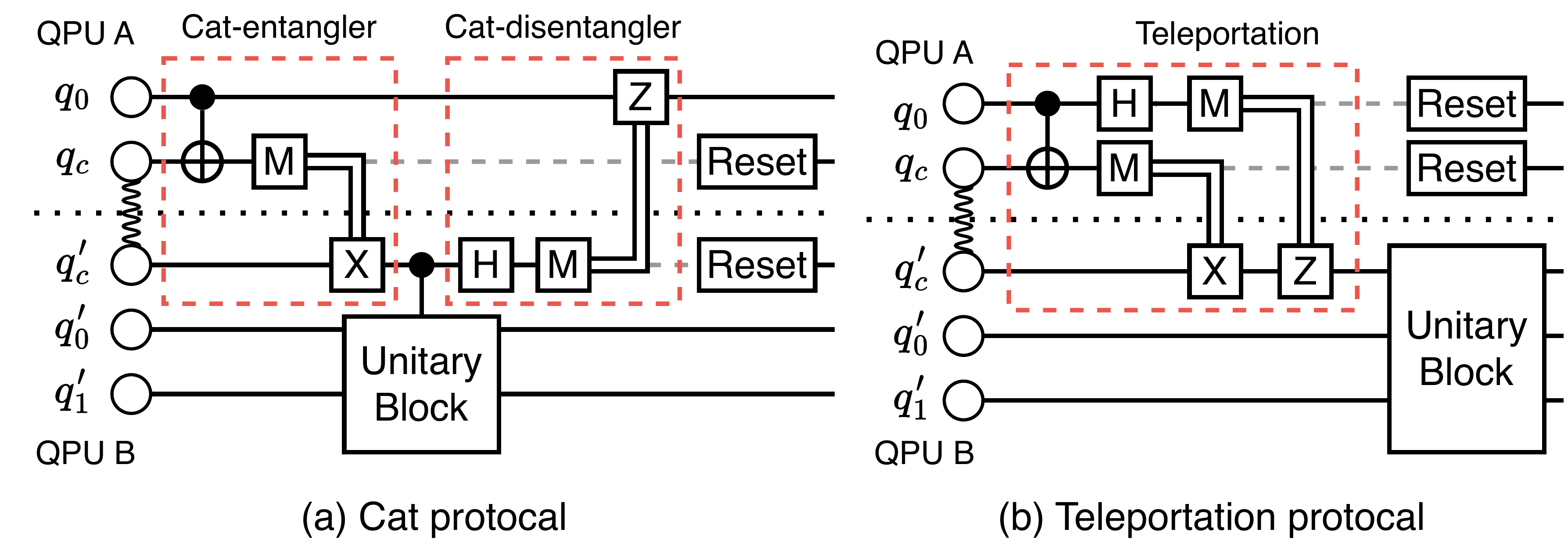}
        
        \caption{(a) Cat protocol and (b) TP protocol for realizing inter-QPU gates with inter-QPU EPR pairs.}
        \label{fig:CAT_TP}
\end{figure}

\textbf{Quantum communication protocols}
Communication between different QPUs can be achieved through EPR pairs (wavy lines in Fig.~\ref{fig:CAT_TP}) established between the QPUs with different protocols.
The Cat protocol \cite{cat-comm} (Fig.~\ref{fig:CAT_TP}(a)) can realize a block of control gates sharing the same control qubit without transferring data from a QPU to another. In contrast, the TP protocol (Fig.~\ref{fig:CAT_TP}(b)) allows any pattern of remote gates by transferring a data qubit from a QPU to another through teleportation (from $q_0$ to $q'_c$ in Fig.~\ref{fig:CAT_TP}(b)).

\textbf{DQC compilation}
Over the last few years, various compiler optimizations have been developed 
for DQC. Some previous work optimizes 
\cite{andres2019automated,baker2020time,beals2013efficient,dadkhah2022reordering,daei2020optimized,davarzani2020dynamic,zomorodi2018optimizing} 
qubit placement without considering transformation or routing strategies to reduce inter-QPU communications, while other work \cite{diadamo2021distributed,ferrari2021compiler,haner2021distributed,AutoComm,QuComm} 
optimizes communications when routing program qubits among QPUs. These approaches are orthogonal to our compiler and can be combined with our optimizations. Some compilers \cite{peng2020simulating,tang2021cutqc} 
have also been proposed to implement DQC without inter-QPU communications by cutting large circuits into smaller ones and running them on different devices, yet they require non-scalable classical post-processing.

Furthermore, recent work \cite{kim2024fault} has proposed an efficient implementation of surface code for DQC. Our work can be considered as a higher-level optimization than it, as each of our qubit can be considered as either a physical qubit or a logical qubit. Combination with QEC implementations leads to repeated generation of multiple EPR pairs between the same pair of QPUs. 
This is effectively equivalent to either a reduced network bandwidth (if generated in parallel) or an increased EPR generation latency (if generated sequentially), or both (if hybrid).

\textbf{Long-rang quantum networks}
Long-range quantum communication relies on repeaters to mitigate significant photon loss over extended distances. Establishing end-to-end entanglement links requires specialized quantum devices and memory qubits at each repeater, coordinated through complex protocols for probabilistic entanglement swapping. Current EPR distribution protocols~\cite{pant2019routing, shi2020concurrent, zhao2021redundant, zhao2022e2e, li2022fidelity} are designed to maximize EPR throughput by strategically selecting successful links among memory qubits. However, these protocols are tailored for repeater networks and reduce to simple shortest-path searches in memoryless QDCs. As a result, they can not address the unique challenges of the QDC architecture effectively.

\subsection{Quantum Data Center (QDC)}
\begin{figure}[tp]
    \centering
    \includegraphics[width=\linewidth]{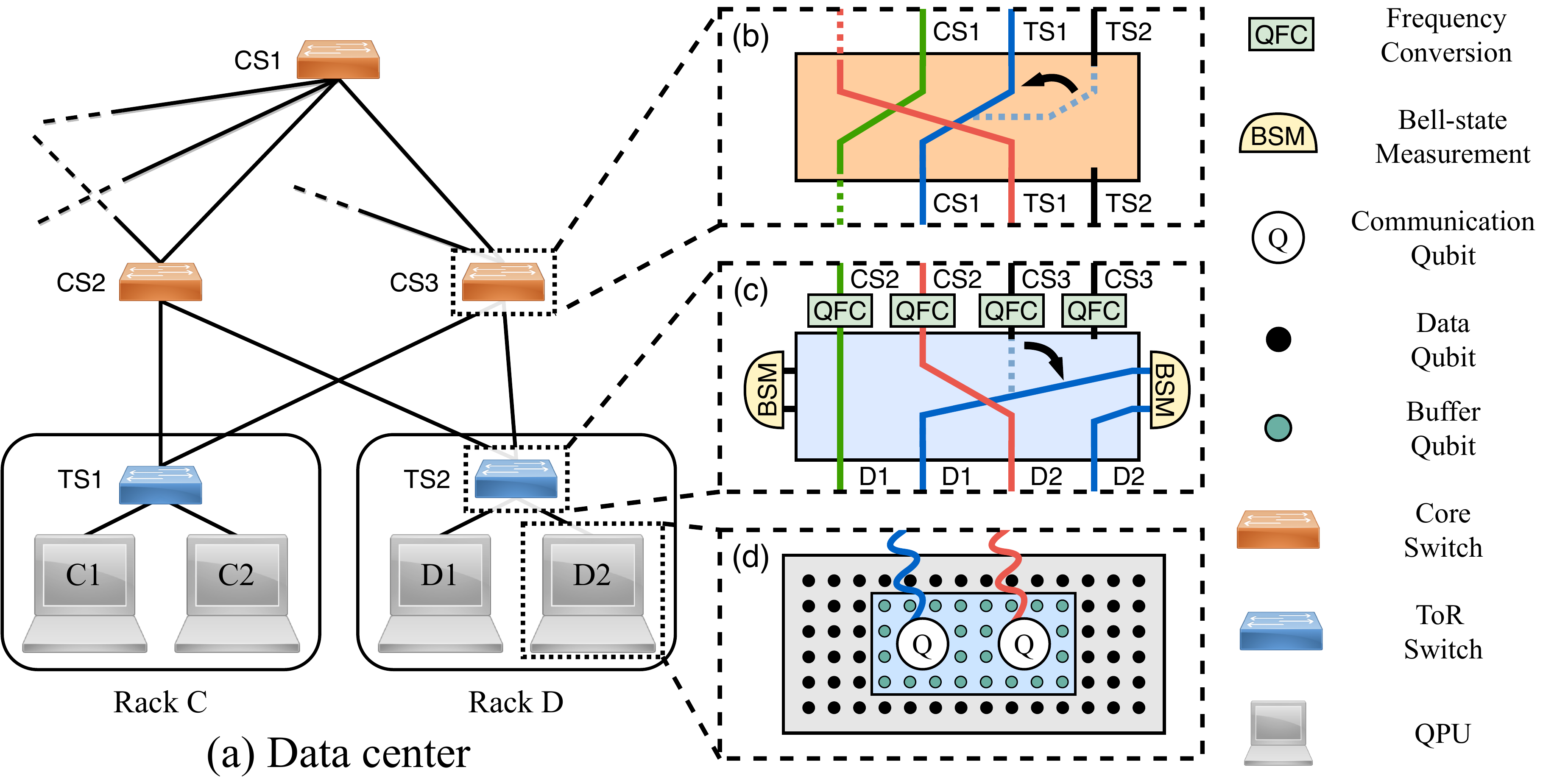}
    \caption{(a) A QDC network with (b) classical inter-rack switches and (c) ToR switches equipped with a QFC on each outward port and some BSMs. (d) Each QPU has two dedicated communication qubits and many computation qubits.}
    
    \label{fig:network_model}
\end{figure}

\textbf{QDC networks} As illustrated by Fig.~\ref{fig:network_model}, a QDC connects QPUs through a local optical network consisting of optical fibers and reconfigurable optical swtiches.
Edges in the network present possible physical channels connecting nodes (i.e., switches and QPUs) through optical fibers. Channel capacity can be increased via multiplexing, e.g., multiple optical fibers or multiple wavelengths through a single fiber, represented by an edge weight $w > 1$.
For instance, Fig.~\ref{fig:network_model}(b)(c) demonstrate a case of edge weight 2 where each pair of nodes are connected by 2 fibers.

\textbf{Optical switch}
The reconfiguration latency of switches ranges from multiple of 100 ns to 100 ms, with the latency increasing with the number of ports~\cite{stepanovsky2019comparative}. Moreover, a faster switch reconfiguration also leads to an increased photon loss.
For switches with about {$16\times 16$} ports \todoo{(or 32 ports)}, the reconfiguration latency is typically {$\sim $ 1 ms} with a photon loss rate around 1 dB~\cite{polatis}, while the latency of larger switches $1024\times 1024$ ports commercially available are typically around 100 - 200ms. In this paper, we will adopt the value of $1$ ms if not stated otherwise. 
\todoo{Each ToR can be equipped with BSM devices to facilitate EPR pair generation. These BSMs can be shared among QPUs in the same rack by ToR reconfiguration, as shown in Fig.~\ref{fig:network_model}(c).}

\begin{figure}[h!]
        \centering
        \includegraphics[width=\linewidth]{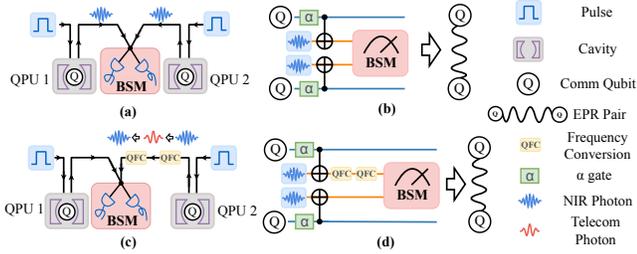}
        \caption{In-rack EPR pair generation: physical process (a) and corresponding circuit (b). Cross-rack EPR pair generation: physical process (c) and corresponding circuit (d)}
        
        \label{fig:EPR_gen}
\end{figure}

\textbf{EPR pair generation}
As depicted in Fig.~\ref{fig:network_model}(d), the generation of EPR pairs requires dedicated communication qubits on each QPU with spin-photon interface between the stationary communication qubits and flying photonic qubits. Fig.~\ref{fig:EPR_gen} illustrates the EPR generation protocol~\cite{monroe2014large,dirk_tutorial}, where optical switches are eliminated for simplicity.

\begin{figure*}[htp]
    \centering
    \includegraphics[width=\textwidth]{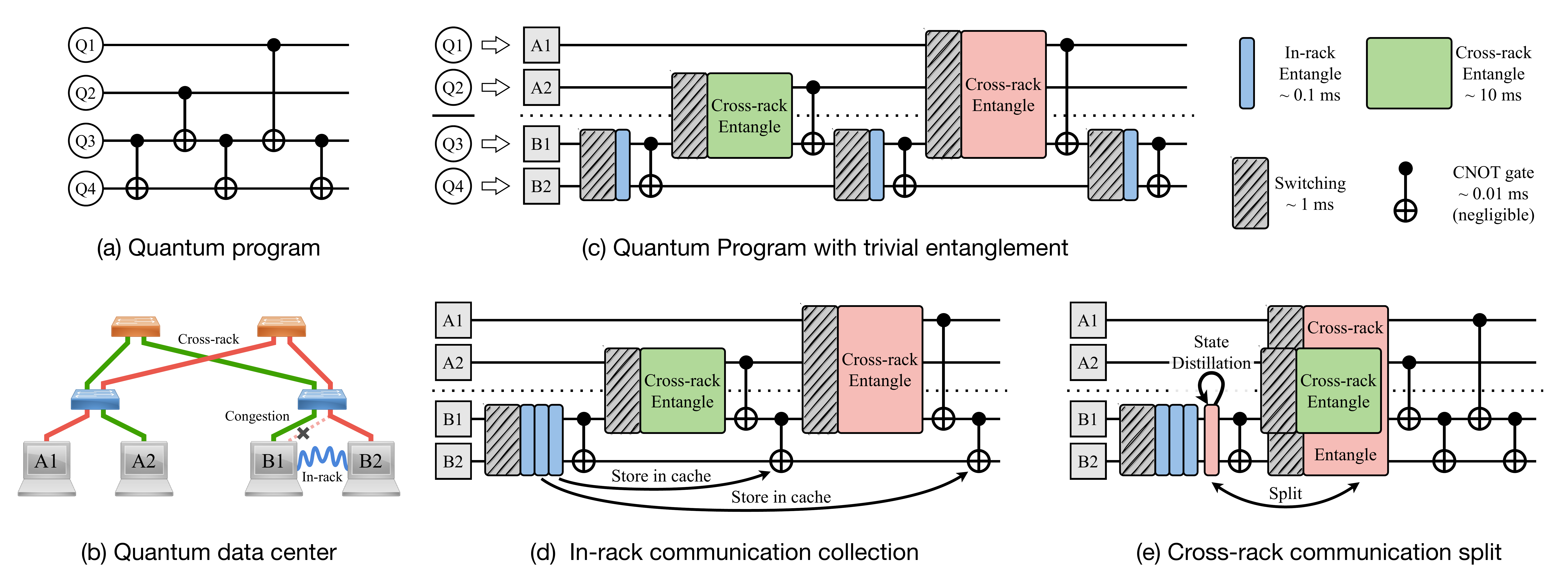}
    \caption{Deployment of a quantum program (a) on to a quantum data center (b). The in-rack communication time is reduced from 3.3 ms (c) to 1.3 ms (d) by a collection. The overall communication time is reduced from 23.3 ms (d) to 12.4 ms (e) by splitting a congested cross-rack communication into a non-congested cross-rack communication and an in-rack communication, where the in-rack one can be distilled at a low cost to enhance its fidelity.}
    \label{fig:motivation}
\end{figure*}

For in-rack communication, as shown Fig.~\ref{fig:EPR_gen}(a), 
communication qubits are prepared in the superposition state $\sqrt{\alpha} \ket{\uparrow}+\sqrt{1-\alpha}\ket{\downarrow}$ and then driven the spin to emit a photon via spontaneous emission, leading to an entangled spin-photon state. Next, the two photonic qubits are directed towards a beam splitter and are eventually measured by the two single-photon detectors. We only post-select the single detection events to project the spin-spin state into $\ket{\phi_+} = (\ket{\uparrow\downarrow}+\ket{\uparrow\downarrow})/\sqrt{2}$. This effectively implements a Bell-state measurement (BSM) with $2\alpha(1-\alpha)$ success probability. Fig.~\ref{fig:EPR_gen}(b) shows the equivalent quantum cirtuit, where the emission process effectively acts as a CNOT gate between the spin and photonic qubit. For cross-rack communication, the protocol is similar but requires quantum frequency converters, as shown in Fig.~\ref{fig:EPR_gen}(c)(d). This is because our ToR switches and BSM devices operate at the near-infrared (NIR) where commercially available optical fiber or switches are not optimized (in terms of the photon loss rate).  To extend the communication range to other racks, we convert the NIR photon into telecom regime and back via bidirectional quantum frequency converters~\cite{van2022entangling,saha2023low,knaut2024entanglement,bersin2024telecom,bersin2024development}.

\textbf{EPR rate and fidelity}
We now estimate the rates and fidelity for in-rack and cross-rack EPR pair generation with some hardware parameters available from the literature.
We note that our EPR generation process is probabilistic due to the probabilistic BSM and due to photon loss during transmission from spin to fiber and as they travel through the switches. 
Hence, the success probability is found to be $p = 2 \alpha \eta$ where $\alpha$ is the initial state parameter and $\eta \ll 1$ denotes the overall photon transmission rate (i.e., overall photon loss rate is $1-\eta$). As a result, we consider a repeat-until-success protocol which keeps trying to generate an EPR pair until getting a positive signal in the BSM. 
Let $\tau_0$ be the operation time for each attempt, then the average time for a successful EPR pair generation is ${\tau}=\tau_0/p$. Considering hardware parameters~\cite{monroe2014large,saha2023low,zhou2024long,stephenson2020high} $\alpha=0.05$, $\eta=0.1$ (i.e., 10 dB loss), and $\tau_0^{-1}\sim 1$ MHz, we obtain {$\tau_\text{ToR}= 0.1$} ms for in-rack EPR pair generation. For the inter-rack communication, the EPR generation rate is reduced by a factor of $100$ (i.e., $20$ dB additional loss) as the transmission rate is further reduced due to the signal attenuation in the second NIR switch, and two QFC devices, leading to $\tau_\text{inter}=10$ ms.

For EPR fidelity, because of false positive signals, we obtain a noisy Bell state in the form of $\hat \rho = (1-\alpha) \ket{\phi_+}\bra{\phi_+}+ \alpha \ket{\uparrow\uparrow}\bra{\uparrow\uparrow}$, and the resulting fidelity is then given by $F = 1-\alpha$. 
With $\alpha=0.05$,
we obtain $F_\text{ToR} = 0.95$ for in-rack EPR pair generation. We further assume that the fidelity is dropped by another $10\%$ due to conversion infidelity in QFCs~\cite{knaut2024entanglement,zhou2024long,stephenson2020high}, resulting in a lower fidelity $F_\text{inter}=0.85$ for cross-rack EPR pair generation.

\section{Motivation}
\label{sect:motivation}

To overcome the unique challenges of QDCs, we propose a novel compiler that hides latencies of cross-rack communications with a minimized number of additional in-rack communications. This section introduces the optimizations in our compiler with a motivating example, including collective generation of in-rack EPR pairs and parallelized generation of cross-rack EPR pairs. 
To facilitate these two optimizations, techniques for resolving deadlock and buffer congestion are also required, which will be introduced with more technical details in the next Section.

Fig.~\ref{fig:motivation}(c) demonstrates an example of scheduling EPR pair generations without decoupling them from communications. It deploys a circuit in Fig.~\ref{fig:motivation}(a) to the QDC in Fig.~\ref{fig:motivation}(b), scheduling EPR pair generations (depicted as rectangles in Fig.~\ref{fig:motivation}(c)) right before they are needed by inter-QPU CNOT gates, with switch reconfiguration represented by the shaded rectangles in Fig.~\ref{fig:motivation}(c). Specifically, each line in Fig.~\ref{fig:motivation}(c) represents a qubit on a different QPU, with the three gates between QPU $B_1$ and $B_2$ requiring in-rack EPR pairs and the other two gates requiring cross-rack EPR pairs according to Fig.~\ref{fig:motivation}(b). Adopting latencies of in\_rack = 0.1 ms, reconfig = 1 ms and cross\_rack = 10 ms, neglecting the much shorter gate execution time, the execution of the 5 remote gates requires 25.3 ms in total. As will be seen, this can be reduced to 12.4 ms with the following optimizations (Fig.~\ref{fig:motivation}(d)(e)).

The first optimization is the collective generation of near future in-rack EPR pairs, which reduces their average latency by avoiding frequent switch reconfigurations. Through a look-ahead into the program, the compiler finds that three in-rack pairs between QPU $B_1$ and $B_2$ are needed in the near future (blue rectangles), which takes 0.3 ms only. However, the switch reconfigurations for enabling the optical channel for three times takes 3 ms (shaded rectangles before the blue ones).
To reduce this overhead, it schedules the generations of these three in-rack EPR pairs collectively, as shown by the consecutive blue rectangles in Fig.~\ref{fig:motivation}(d).
These generated EPR pairs are then stored temporarily in the buffer before usage.
With this collection, the time for generating the three in-rack EPR pairs is reduced from 3.3 ms to 1.3 ms.

The second optimization is the parallelization of near future cross-rack EPR pairs, which maximizes the utilization of network bandwidth by splitting congested EPR pairs into non-congested ones.
In Fig.~\ref{fig:motivation}(d), generations of the two cross-rack EPR pair $(A_2,B_1)$ and $(A_1,B_1)$ are sequential, as QPU $B_1$ can communicate with only one other QPU at a time (assuming edge weight $=1$). That is, in Fig.~\ref{fig:motivation}(b), the green path conflicts with the red dashed path.
To parallelize them, we can split the congested EPR pair $(A_1,B_1)$ by allowing $B_1$ to borrow bandwidth from $B_2$, a QPU in the same rack with $B_1$.
Specifically, the cross-rack EPR pair $(A_1, B_1)$ is split into a new cross-rack EPR pair $(A_1, B_2)$ and an additional in-rack EPR pair $(B_1, B_2)$, followed by an entanglement swapping between these two pairs once they are both prepared. In this way, we can generate the new cross-rack EPR pair $(A_1, B_2)$ in parallel with the originally conflicted $(A_2, B_1)$ and store it in the buffer, while scheduling an additional in-rack EPR pair $(B_1, B_2)$ in the collective generation, as shown in Fig.~\ref{fig:motivation}(d). 
This further reduces the overall latency from 23.3 ms (Fig.~\ref{fig:motivation}(d)) to 12.4 ms (Fig.~\ref{fig:motivation}(e)).

As the split of cross-rack communicates incurs additional in-rack EPR pairs, it poses a risk of reducing the overall fidelity of quantum computing. To mitigate this, we enhance the fidelity of these additional in-rack EPR pairs by scheduling multiple copies of each and implementing entanglement distillation among them, as shown in Fig.~\ref{fig:motivation}(e). 
This can be achieved with a low cost as the multiple copies can be scheduled collectively with other in-rack EPR pairs. Given that in-rack EPR pairs have a 95\% fidelity,
a distillation by two copies results in an EPR pair of $>$ 96.5\% fidelity (where we approximated our input states as Werner states~\cite{bennett1996purification}) with 93.6\% success probability. This fidelity can be further enhanced by using more copies. 
Note that cross-rack EPR pairs and original (non-split) in-rack EPR pairs can also be distilled upon requests. This can be easily accommodated by our framework, which is equivalent to an increased latency of EPR pair generation.

\section{Framework Design}

\subsection{Preprocessing}
Our compiler can be used in combination with previous compilers \cite{AutoComm,QuComm} that reduce the required EPR pairs. Considering the dynamic reconfigurability of QDCs, these compilers for static network topology should be applied by assuming a full connection between QPUs. 
Output of these compilers is a list of required EPR pairs, with corresponding QPUs and communication protocols specified (i.e., Cat vs. TP, see section~\ref{sect:background}).

We convert this output to a directed acyclic graph (DAG) by imposing dependencies among them, with each node in DAG representing an EPR pair, and each directed edge representing a dependency. Specifically, we consider a later EPR pair in the output list as dependent on an earlier one if their involved QPUs overlap. Note that this imposed dependency can deviate from the real dependency. From the perspective of program, this is an overly loose standard as some EPR pairs may have hidden dependency induced by bandwidth contention. However, this allows for a more flexible scheduling of EPR pair generations, with issues of the loose standard being fixable by an efficient retry mechanism that will be explained later. From the perspective of hardware constraints, this is also an overly strict standard since the overlapped QPUs may have multiple communication qubits which are able to work in parallel. However, this is enough for subsequent processing as our scheduling will automatically parallelize the generation of EPR pairs if they are indeed independent.

\begin{figure*}[pth]
        \centering
        \includegraphics[width=0.87\linewidth]{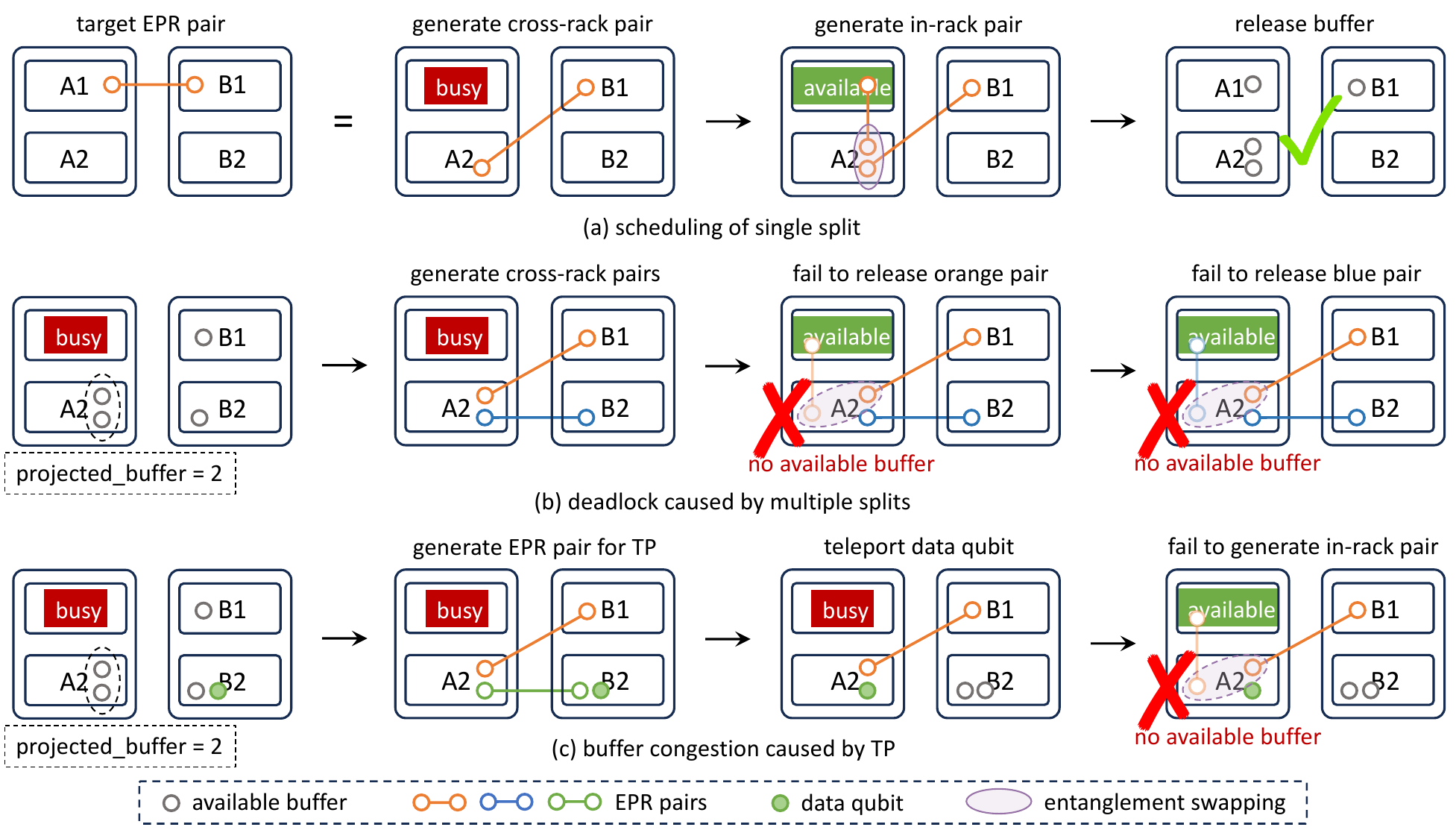}
        
        \caption{(a) A normal EPR split. (b) Deadlock caused by multiple EPR splits. (c) Buffer congestion caused by TP communication.}
        \label{fig:split_congestion}
\end{figure*}

\subsection{Scheduling with In-rack Collection} 
This subsection introduces the scheduling principle in the presence of collective in-rack EPR pair generation, temporarily without cross-rack communication split.
Due to limitations in communication qubits and network bandwidth, at each time slice it depends on four conditions to decide whether an EPR between two QPUs can be scheduled, as listed below. With our collective EPR pair generation and program-aware look-ahead scheduling, these conditions vary with whether an EPR pair is in-rack or cross-rack and whether it is in the front layer of the DAG.

\begin{enumerate}
    \item Available communication qubits on both QPUs
    \item Available BSMs on the rack of either QPU
    \item Available optical channel in the network 
    \item buffer\_size + avail\_comm $\ge$ threshold $\cdot$ not\_in\_front \\(threshold $\ge$ \#comm\_qubits per QPU for both QPUs) 
    
\end{enumerate}

\textbf{In-rack vs. cross-rack} The first three are mandatory requirements unless the EPR pair to be generated is an in-rack one which can be combined with previous in-rack EPR pairs. That is, given an EPR pair between QPU $A$ and $B$, we first check if there are available communication qubits on both $A$ and $B$, and check whether there is an available BSM on the ToR switch of either $A$'s rack or $B$'s rack. 
Then through a shortest path search, we check if there is an available path in the network between $A$ and $B$ that is not occupied by any other communication. For a cross-rack communication, failure to satisfy any of these three conditions implies that the EPR pair cannot be scheduled at the current time. For an in-rack communication, if it fails to satisfy these three conditions, we further check if the failure is caused by bandwidth occupancy of another in-rack EPR pair generation between the same QPUs $A$ and $B$. If so, we combine this EPR pair with those between $A$ and $B$ by scheduling it right after them. 

\textbf{Front layer vs. non-front layer} The fourth condition is not always mandatory, depending on
the position of the EPR pair in the DAG.  
For EPR pairs in the front layer, we allow their generations as long as the three conditions above are satisfied, as they are required by the program urgently. However, for those not in the front layer, which are less urgent, we set this more strict condition to mitigate buffer congestion, which requires the total number of available buffer qubits and communication qubits on each involved QPU to exceed a threshold.

\subsection{Scheduling with cross-rack split}
This subsection introduces the modified scheduling principle in the presence of cross-rack EPR split. 
As illustrated by Fig.~\ref{fig:split_congestion}(a), the split of a cross-rack EPR pair $(A_1,B_1)$ results in a new cross-rack pair $(A_2,B_2)$ and an additional in-rack pair $(A_1,A_2)$. This is followed by an entanglement swapping that merges them into the required pair $(A_1,B_1)$ and a buffer release after this EPR pair is consumed by communication. This can help maximize the utilization of network bandwidth since it allows a busy QPU $A_1$ to borrow communication qubits of another QPU $A_2$ in the same rack and share its buffer. However, its flexibility also brings a risk of deadlocks or buffer congestion if available buffer sizes on involved QPUs are not large enough. As a result, we impose the following buffer conditions to mitigate this risk, then introduce the modified EPR pair scheduling principle in the presence of cross-rack split.

\textbf{Conditions for split}
To allow EPR spilt, the first condition is to ensure there are available communication qubits on another QPU in the same rack as the busy QPU, as listed below. Then, we make sure there is enough buffer on the involved QPUs by checking the second condition below. 

\begin{enumerate}
    \item Available communication qubits on another QPU in the same rack as the busy QPU
    \item $\mathrm{projected\_buffer} - \mathrm{reserved\_buffer} \ge m_{QPU}$ for each QPU involved in the post-split EPR pairs
\end{enumerate}

We first define $\mathrm{projected\_buffer}$ of a QPU as the buffer size it would have if all currently scheduled EPR pairs were consumed by the execution of corresponding communications. This reflects the maximum buffer size available in the near term.
When calculating this variable, if an EPR pair is scheduled for a Cat protocol, the buffer size of each involved QPU should increase by 1 after the communication, as the two buffer qubits occupied by this EPR pair on the two involved QPUs would be released. In contrast, if the EPR pair is scheduled for a TP protocol with data teleported from QPU $A$ to QPU $B$, then the buffer size of $A$ should increase by 2 after the communication, while the buffer size of $B$ should remain unchanged.

With this variable, a basic condition for EPR split is that the $\mathrm{projected\_buffer}$ on each QPU should at least be able to accommodate all EPR pairs induced by the split, i.e., $\mathrm{projected\_buffer} \ge m_{QPU}$, with $m_{QPU}$ being the number of post-split EPR pairs on each involved QPU. For example, if an EPR pair $(A,B)$ is split into $(A,A')$ and $(A',B)$, then there should be $m_A=m_B=1$ and $m_{A'}=2$, as each of $A$ and $B$ needs to store only one EPR pair, while $A'$ needs to store two EPR pairs.

However, this basic condition is not sufficient to prevent deadlock caused by multiple EPR splits. We illustrate this deadlock risk with an example in Fig.~\ref{fig:split_congestion}(b).
The two orange EPR pairs $(A_1, A_2)$ and $(A_2, B_1)$ are the post-split pairs of an EPR pair $(A_1, B_1)$, while the two blue EPR pairs $(A_1, A_2)$ and $(A_2, B_2)$ are the post-split pairs of an EPR pair $(A_1,B_2)$. Supposing QPU $A_2$ has $\mathrm{projected\_buffer} = 2$, while the two post-split cross-rack pairs are scheduled to be generated simultaneously, then the post-split in-rack EPR pairs would never have a chance to be scheduled due to a deadlock. This is because the orange in-rack pair  will be waiting for buffer release from the blue cross-rack pair , which requires the blue in-rack pair  to be generated first; while the blue in-rack pair  will be waiting for buffer release from the orange cross-rack pair , which in turn requires the orange in-rack pair  to be generated first.

To accommodate simultaneous splits of multiple EPR pairs, we need to prevent this deadlock by reserving enough buffer size on QPUs. Specifically, for each EPR split, we reserve $m_\mathrm{QPU}$ buffer qubits from $\mathrm{projected\_buffer}$ until the post-split EPR pairs are all scheduled, with $m_\mathrm{QPU}$ being the number of post-split EPR pairs on each involved
QPU as explained earlier. That is, we track a variable $\mathrm{reserved\_buffer}$ on each QPU, which is initiated as 0. For each split, we increase the $\mathrm{reserved\_buffer}$ of each QPU by $m_{QPU}$ when the first post-split EPR pair of this split is scheduled, then decrease them by $m_\mathrm{QPU}$ once all post-split pairs of this split are scheduled. With this reservation, the condition for an EPR split becomes  $\mathrm{projected\_buffer} - \mathrm{reserved\_buffer} \ge m_{QPU}$.

\textbf{Modified scheduling principles}
The post-split EPR pairs can be scheduled in the same way as regular pairs by updating the DAG with the post-split pairs. However, their presence
can cause more buffer congestion, especially when other EPR pairs need to be scheduled for TP protocols.
For example, in Fig.~\ref{fig:split_congestion}(c), the green EPR pair is for a TP communication that teleports a qubit from $B_2$ to $A_2$. This teleported data would release a buffer on $B_2$, but not on $A_2$, as the teleported data would cancel the released buffer on $A_2$. 
As a result, the buffer size $\mathrm{projected\_buffer} = 2$ of $A_2$ becomes insufficient even if was originally enough to accommodate the two orange pairs in Fig.~\ref{fig:split_congestion}(a).

To address this, we modify the scheduling principle as the following, distinguishing Cat and TP protocols while taking the buffer reservation into account.

\begin{enumerate}
    \item Available communication qubits on both QPUs
    \item Available BSMs on the rack of either QPU
    \item Available optical channel in the network
    \item For Cat protocol or the source QPU of a TP protocol:\\$\mathrm{projected\_buffer} - \mathrm{reserved\_buffer} \\+ \mathrm{avail\_comm} > \mathrm{threshold}\cdot\mathrm{not\_in\_front\_layer}$ \\
    (threshold $\ge$ \#comm\_qubits per QPU for both QPUs)
    \item For the destination QPU of a TP protocol:\\
    $\mathrm{projected\_buffer} - \mathrm{reserved\_buffer} \\+ \mathrm{avail\_comm} - 1 \ge \mathrm{threshold}\cdot\mathrm{not\_in\_front\_layer}$ \\
    (threshold $\ge$ \#comm\_qubits per QPU for both QPUs)
    
\end{enumerate}

The first three conditions are the same with previous. The fourth condition is still an optional condition that only applies to EPR pairs in non-front layers of the DAG. It is modified from the original fourth condition by substituting the available buffer size with  $\mathrm{projected\_buffer} - \mathrm{reserved\_buffer}$. 
The fifth condition puts a more strict constraint on EPR pairs for TP protocol, considering that a teleportation will occupy a buffer qubit on its destination QPU. That is, for each EPR pair for TP protocol (either original or post-split), an additional buffer qubit needs to be reserved on the destination QPU to accommodate the teleported data qubit.
In contrast, this condition is not required for Cat protocol, because even if one of the reserved qubit is temporarily occupied by the EPR pair of a Cat protocol, it will be released once the communication completes.

\subsection{Post-split distillation}
Since the split of cross-rack communications requires additional in-rack EPR paris, it reduces the overall fidelity since the fidelities of EPR pairs are much lower than the local gates on each QPU. To improve the overall fidelity, we incorporate an entanglement distillation for the post-split in-rack EPR pairs. Given the 95\% fidelity of in-rack EPR pairs, 
a distillation of two EPR pairs enhances the fidelity to \todoo{$> 96.5\%$}.
However, the additional EPR pair required by the distillation also needs to occupy a buffer qubit on the QPU. As a result, the reserved buffer in the conditions above should become $m=3$. This reservation does not need to increase with the number of EPR pairs $k$ used for each entanglement distillation, because the $k-1$ sacrificed EPR pairs can reuse the same buffer qubit sequentially.

\subsection{Algorithm}

\textbf{Look-ahead scheduling} 
At each time slice $t_0$, our compiler looks into the subgraph of the first $l$ layers of the DAG, maximizing the number of scheduled EPR pairs greedily. This scheduling consists of two rounds, with the first round for scheduling of regular EPR pairs and the second round for EPR split and the scheduling of post-split EPR pairs. 

In the first round, it tries scheduling each node in the subgraph in the ascending order of their layers. If the EPR pair represented by this node satisfies the conditions in Summary 3, then we schedule it, removing the node and updating the dependencies between the remaining nodes in the subgraph. Otherwise, we continue with the next node until no EPR pair in the first $l$ layers satisfies the conditions.

Then we conduct a second round of scheduling if there is remaining network bandwidth at $t_0$. Specifically, we try splitting each node in the remaining subgraph in the ascending order of node layer. If the EPR pair satisfies the conditions in Summary 2, we split the EPR pair, scheduling the post-split cross rack pair and adding the post-split in-rack pairs into the subgraph. Then we continue to the next node in the updated subgraph until all nodes in the subgraph are traversed. After that, we repeat these two round for time $t_0+1$.

\textbf{Auto retry} 
While the conditions for EPR split and scheduling can significantly mitigate deadlock and buffer congestion, there is still possibility that they may occasionally occur due to the flexibility in scheduling. We illustrate this with two simple examples and then address it with a retry-mechanism for scheduling.

To address this challenge, we implement an automatic retry mechanism that reverts to a saved state and retries scheduling with a more conservative strategy if a deadlock or congestion occurs. The most conservative approach would be to follow the exact order of EPR pair generations set by the pre-processing stage. Although this ensures progress, it significantly limits the opportunities to parallelize cross-rack communications. Therefore, we propose a medium conservative strategy that allows parallelization of communications between non-overlapping QPU pairs.

Specifically, this strategy checks the list of EPR pairs given by the pre-processing in their strict order, treating an EPR pair, say between $A$ and $B$, as generatable if either previous pairs do not involve $A$ and $B$ or they are also generatable.
Then these generatable EPR pairs will be scheduled if the network has bandwidth.
Note that this strategy does not guarantee a congestion-free communication like the most conservative strategy.
Whenever this medium conservative strategy leads to a deadlock or congestion, we revert to a previous saved status again and applies the most conservative strategy as a fallback. 

\section{Evaluation}\label{sect:eval}

\subsection{Experiment Setup}
\label{sect:setup}

\textbf{Architecture setups} 
In the primary experiment, we evaluate our framework on the CLOS network architecture \cite{CLOS,tate2013ibm} as shown in Fig.~\ref{fig:intro}, with 25\% of the total computation qubits reserved as buffer, as listed in Table~\ref{tab:setting_table}. 
Each QPU is assumed to have two communication qubits. To allow all communication qubits in a rack to work in parallel, we assume each ToR switch to have \#BSMs = $2\times$ \#QPUs / rack and each switch to be a \#BSMs $\times$ \#BSMs switch (or 2$\times$ \#BSMs ports), 
We take a look-ahead depth of \todo{10}, adopting 0.1ms, 1ms and 10ms as the latencies of in-rack EPR pair generation, switch configuration and cross-rack EPR pair generation respectively, according to Section~\ref{sect:background}.
The settings of experiments are listed in Table~\ref{tab:setting_table}.
These parameters will be further varied in subsequent experiments.

\begin{table}[h]
    \centering
    \caption{\todoo{Program and architecture settings}}
    \resizebox{0.48\textwidth}{!}{
        \renewcommand*{\arraystretch}{1}
        \begin{normalsize}
            \begin{tabular}{|p{2cm}|p{0.8cm}|p{0.8cm}|p{1.7cm}|p{1.4cm}|p{1.7cm}|}
        
        \hline
        Benchmark & \#rack & \#QPUs / rack & \#Data Qubits / QPU  &  Buffer Size / QPU & Comm Qubit / QPU \\
        \hline
        
         program-480 & 4 & 4 & 30 & 10 & 2 \\ \hline
         program-608 & 4 & 4 & 38 & 12 & 2 \\ \hline
         program-720 & 4 & 4 & 45 & 15 & 2 \\ \hline

         program-360 & 4 & 3 & 30 & 10 & 2 \\ \hline
         program-480 & 4 & 4 & 30 & 10 & 2 \\ \hline
         program-600 & 4 & 5 & 30 & 10 & 2 \\ \hline
         program-720$^*$ & 4 & 6 & 30 & 10 & 2 \\ \hline

         program-240 & 4 & 3 & 20 & 7 & 2 \\ \hline
         program-540 & 9 & 3 & 20 & 7 & 2 \\ \hline
         program-960 & 16 & 3 & 20 & 7 & 2 \\ \hline
         spine-leaf-720 & 6 & 4 & 30 & 10 & 2 \\ \hline
         fat-tree-960 & 8 & 4 & 30 & 10 & 2 \\ \hline
                
    \end{tabular}
    \end{normalsize}
    }
    \label{tab:setting_table}
\end{table}

\textbf{Benchmark programs} 
We select a set of benchmark programs including building blocks of quantum applications and practical quantum algorithms: multi-control target gate (MCT) \cite{MCT},
Quantum Fourier transform (QFT) \cite{QFT},
Grover's Algorithm (Grover) \cite{grover_impl}, 
Ripple-Carry Adder (RCA)~\cite{RCA}. For Grover, we consider the secret string with all ones and repeat the iteration by \todo{100} times. Moreover, we increase the complexity of the RCA circuit by repeating the adder for \todo{100} iterations, which effectively adapts it to a sum calculation.

\textbf{Metrics} 
We evaluate the compiler performance with three metrics. The first metric is the overall communication latency, normalized by the latency of switch reconfiguration. We ignore the computation time within each QPU as it is much faster than inter-QPU communication. The second and the third metrics together measure the fidelity overhead. Specifically, the second metric is the number of EPR pairs used for communication, weighted by the infidelity of each pair, denoted as \emph{\#EPR}. According to the 15\% and 5\% infidelity of cross-rack and in-rack EPR pairs (see Section~\ref{sect:background}), we count them by weights 1 and 0.33, respectively. For the distilled in-rack EPR pairs, we count each of them with a weight \todoo{0.23} based on their \todoo{$<3.5\%$} infidelity. The third metric is the average wait time of EPR pairs in buffer, which is also normalized by reconfiguration latency. We list these two overheads separately as the effect of wait time depends on the coherence time of computation qubits, which is not decided by the network but varies with different QPU technology.

\begin{table*}[tp]
    \centering
    \caption{The performance of our compiler and its comparison to the baseline. Units of latency and wait time are the latency of switch reconfiguration. The number of EPR pairs is weighted by infidelity of each EPR pair.}
    \resizebox{0.98\textwidth}{!}{
        \renewcommand*{\arraystretch}{1.03}
        \begin{small}
            \begin{tabular}{|p{0.3cm}|p{1.8cm}|p{1.4cm}|p{1.4cm}|p{1.2cm}|p{1.4cm}|p{1.2cm}|p{1.2cm}|p{1.2cm}|p{1.2cm}|p{1.2cm}|}
        
        \hline
        Experiment & Benchmark & Baseline:$\quad\quad$ Latency & Ours:$\quad\quad$ Latency & Improv. Factor & Baseline: \#EPR & Ours: \#EPR & EPR Overhead & Baseline: wait time& Ours: wait time & Additional wait time  \\
        \hline
        \multicolumn{1}{|c|}{\multirow{12}{*}{\parbox{1.8cm}{ Increase\\ \#qubits/QPU}}}
&MCT-480&2,312&485&$\boldsymbol{4.77\times}$&95.00&98.03&3.09\%&1.98&6.75&4.77\\
\cline{2-11}
&MCT-608&2,312&454&$\boldsymbol{5.09\times}$&95.00&98.50&3.55\%&1.98&6.36&4.39\\
\cline{2-11}
&MCT-720&2,312&382&$\boldsymbol{6.05\times}$&95.00&98.73&3.78\%&1.98&8.23&6.25\\
\cline{2-11}
&QFT-480&121,728&16,693&$\boldsymbol{7.29\times}$&2,070.00&2,334.37&11.32\%&1.30&6.87&5.57\\
\cline{2-11}
&QFT-608&155,960&20,781&$\boldsymbol{7.50\times}$&2,622.00&2,960.80&11.44\%&1.26&6.92&5.66\\
\cline{2-11}
&QFT-720&194,526&24,670&$\boldsymbol{7.89\times}$&3,105.00&3,518.47&11.75\%&1.00&7.77&6.77\\
\cline{2-11}
&Grover-480&156,213&26,943&$\boldsymbol{5.80\times}$&4,200.00&4,649.63&9.67\%&2.08&8.73&6.65\\
\cline{2-11}
&Grover-608&150,702&27,412&$\boldsymbol{5.50\times}$&4,200.00&4,651.03&9.70\%&1.87&9.54&7.67\\
\cline{2-11}
&Grover-720&156,213&25,883&$\boldsymbol{6.04\times}$&4,200.00&4,695.13&10.55\%&2.08&9.14&7.06\\
\cline{2-11}
&RCA-480&92,259&9,169&$\boldsymbol{10.06\times}$&1,407.00&1,554.00&9.46\%&0.03&8.49&8.46\\
\cline{2-11}
&RCA-608&92,259&9,304&$\boldsymbol{9.92\times}$&1,407.00&1,558.67&9.73\%&0.03&8.57&8.54\\
\cline{2-11}
&RCA-720&92,226&9,395&$\boldsymbol{9.82\times}$&1,404.33&1,557.87&9.86\%&0.01&9.78&9.77\\
        \hline

        \multicolumn{1}{|c|}{\multirow{16}{*}{\parbox{1.8cm}{Increase \#QPUs/rack}}}
&MCT-360&1,476&468&$\boldsymbol{3.15\times}$&63.00&66.50&5.26\%&3.03&5.81&2.78\\
\cline{2-11}
&MCT-480&2,312&485&$\boldsymbol{4.77\times}$&95.00&98.03&3.09\%&1.98&6.75&4.77\\
\cline{2-11}
&MCT-600&3,214&634&$\boldsymbol{5.07\times}$&159.00&161.80&1.73\%&1.17&5.30&4.13\\
\cline{2-11}
&MCT-720$^*$&4,413&921&$\boldsymbol{4.79\times}$&223.00&225.57&1.14\%&0.87&5.27&4.40\\
\cline{2-11}
&QFT-360&78,300&13,504&$\boldsymbol{5.80\times}$&1,310.00&1,476.83&11.30\%&1.48&6.27&4.79\\
\cline{2-11}
&QFT-480&121,728&16,693&$\boldsymbol{7.29\times}$&2,070.00&2,334.37&11.32\%&1.30&6.87&5.57\\
\cline{2-11}
&QFT-600&169,831&20,041&$\boldsymbol{8.47\times}$&2,990.00&3,353.77&10.85\%&1.14&7.00&5.86\\
\cline{2-11}
&QFT-720$^*$&216,372&23,362&$\boldsymbol{9.26\times}$&4,070.00&4,516.83&9.89\%&1.20&7.63&6.43\\
\cline{2-11}
&Grover-360&140,813&29,717&$\boldsymbol{4.74\times}$&3,400.00&3,771.93&9.86\%&2.03&9.96&7.93\\
\cline{2-11}
&Grover-480&156,213&26,943&$\boldsymbol{5.80\times}$&4,200.00&4,649.63&9.67\%&2.08&8.73&6.65\\
\cline{2-11}
&Grover-600&171,613&25,438&$\boldsymbol{6.75\times}$&5,000.00&5,479.97&8.76\%&2.08&8.77&6.68\\
\cline{2-11}
&Grover-720$^*$&187,013&24,580&$\boldsymbol{7.61\times}$&5,800.00&6,308.20&8.06\%&2.07&8.85&6.77\\
\cline{2-11}
&RCA-360&83,470&10,127&$\boldsymbol{8.24\times}$&1,139.00&1,262.90&9.81\%&0.03&9.69&9.66\\
\cline{2-11}
&RCA-480&92,259&9,169&$\boldsymbol{10.06\times}$&1,407.00&1,554.00&9.46\%&0.03&8.49&8.46\\
\cline{2-11}
&RCA-600&101,048&8,916&$\boldsymbol{11.33\times}$&1,675.00&1,834.37&8.69\%&0.03&8.68&8.64\\
\cline{2-11}
&RCA-720$^*$&109,837&8,592&$\boldsymbol{12.78\times}$&1,943.00&2,108.67&7.86\%&0.03&8.78&8.75\\
\hline
        \multicolumn{1}{|c|}{\multirow{12}{*}{\parbox{1.8cm}{Increase \\ \#racks}}}
&MCT-240&1,575&490&$\boldsymbol{3.21\times}$&63.00&66.27&4.93\%&2.74&4.66&1.92\\
\cline{2-11}
&MCT-540&6,514&3,069&$\boldsymbol{2.12\times}$&399.00&411.13&2.95\%&3.12&5.14&2.03\\
\cline{2-11}
&MCT-960&15,369&5,415&$\boldsymbol{2.84\times}$&1,023.00&1,062.67&3.73\%&3.21&5.44&2.23\\
\cline{2-11}
&QFT-240&50,195&9,663&$\boldsymbol{5.19\times}$&873.33&964.10&9.41\%&1.61&6.25&4.64\\
\cline{2-11}
&QFT-540&212,522&28,694&$\boldsymbol{7.41\times}$&4,273.33&4,693.80&8.96\%&2.36&6.28&3.93\\
\cline{2-11}
&QFT-960&564,973&58,482&$\boldsymbol{9.66\times}$&13,233.33&14,225.00&6.97\%&3.05&6.61&3.56\\
\cline{2-11}
&Grover-240&147,468&34,265&$\boldsymbol{4.30\times}$&3,400.00&3,691.20&7.89\%&1.00&9.54&8.54\\
\cline{2-11}
&Grover-540&365,312&38,648&$\boldsymbol{9.45\times}$&8,400.00&9,116.57&7.86\%&0.99&8.63&7.65\\
\cline{2-11}
&Grover-960&665,612&39,969&$\boldsymbol{16.65\times}$&15,400.00&16,467.73&6.48\%&0.98&7.99&7.01\\
\cline{2-11}
&RCA-240&83,470&11,050&$\boldsymbol{7.55\times}$&1,139.00&1,239.80&8.13\%&0.03&8.91&8.88\\
\cline{2-11}
&RCA-540&213,523&12,666&$\boldsymbol{16.86\times}$&2,814.00&3,013.03&6.61\%&0.08&7.10&7.02\\
\cline{2-11}
&RCA-960&399,478&13,240&$\boldsymbol{30.17\times}$&5,159.00&5,374.60&4.01\%&0.03&7.03&7.00\\
        \hline
        \multicolumn{1}{|c|}{\multirow{4}{*}{\parbox{1.8cm}{Spine-leaf \\ topology}}}
&MCT-720&4,611&1,008&$\boldsymbol{4.57\times}$&239.00&246.70&3.12\%&2.24&6.60&4.36\\
\cline{2-11}
&QFT-720&249,317&34,657&$\boldsymbol{7.19\times}$&4,590.00&5,002.30&8.24\%&1.77&9.89&8.12\\
\cline{2-11}
&Grover-720&242,013&34,357&$\boldsymbol{7.04\times}$&6,600.00&7,066.90&6.61\%&2.22&11.63&9.42\\
\cline{2-11}
&RCA- 720&149,316&11,418&$\boldsymbol{13.08\times}$&2,211.00&2,369.43&6.69\%&0.03&10.36&10.33\\

        \hline
        \multicolumn{1}{|c|}{\multirow{4}{*}{\parbox{1.8cm}{Fat tree \\ topology}}}
&MCT-960&6,910&2,497&$\boldsymbol{2.77\times}$&383.00&390.00&1.79\%&2.44&5.82&3.38\\
\cline{2-11}
&QFT-960&421,181&49,568&$\boldsymbol{8.50\times}$&8,070.00&8,794.97&8.24\%&1.99&9.40&7.41\\
\cline{2-11}
&Grover-960&327,813&39,193&$\boldsymbol{8.36\times}$&9,000.00&9,564.67&5.90\%&2.28&12.80&10.52\\
\cline{2-11}
&RCA-960&206,373&12,639&$\boldsymbol{16.33\times}$&3,015.00&3,224.07&6.48\%&0.03&10.52&10.49\\
\hline  
                
    \end{tabular}
    \end{small}
    }
    \label{tab:main_table}
\end{table*}

\textbf{Baseline} We construct our baseline by combining state-of-the-art compiler \cite{QuComm} with a shortest-path EPR generation. Specifically, we obtain the required EPR pairs for each benchmark program based on the idea of buffer reservation in \cite{QuComm} and schedule the generation of each EPR pair through a shortest available path in the network right before it is required by an inter-QPU communication. As a slight optimization of this baseline, we adopt the medium conservative scheduling strategy mentioned previously rather than the most conservative one as we did not observe fatal congestion with the medium strategy during experiments.

\subsection{Primary Experiment Result}

\textbf{Outperformance}
Table~\ref{tab:main_table} presents the comparison of our framework with the baseline as the number of qubits per QPU increases, as the number of QPU per rack increases and as the number of racks. Besides CLOS, we also include two other commonly used network topologies, i.e., spine-leaf topology \cite{tate2013ibm} and fat tree topology \cite{choi2023scalable}. 
It can be seen that our compiler significantly reduces the overall latency, with an average improvement factor of \todo{8.02}. 

The results of both the baseline and our compiler stay stable with \#qubits per QPU, as the increase of \#qubits per QPU of primarily affects QPU computation rather than inter-QPU communication. Only QFT is different, since it has a denser communication pattern than other benchmarks. As a result, the improvement of our compiler also stays stable with \#qubits per QPU. 

The improvement of our compiler increases slightly with \#QPUs per rack with some fluctuation. This shows the ability of our compiler to mitigate the bandwidth contention caused by the increased \#QPUs. Moreover, it increases more significantly with \#racks, exhibiting the capability of our compiler of effectively utilizing the increased cross-rack bandwidth. These results demonstrate the scalability of our compiler. Furthermore, the improvement factors of the other two network topologies are at a similar level with CLOS network, which demonstrates the general applicability of our compiler to various network topologies.
\begin{figure*}[tp!]
        \centering
        \includegraphics[width=1\textwidth]{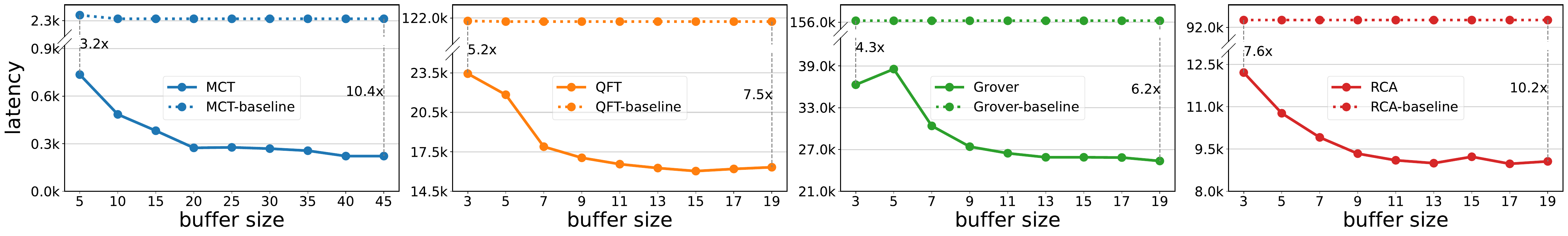}
        \\
        \vspace{-7pt}
        \hspace{20pt}{\footnotesize (a1)}\hspace{115pt}{\footnotesize (a2)}\hspace{115pt}{\footnotesize (a3)}\hspace{115pt}{\footnotesize (a4)}\\
        \vspace{5pt}
        \includegraphics[width=1\textwidth]{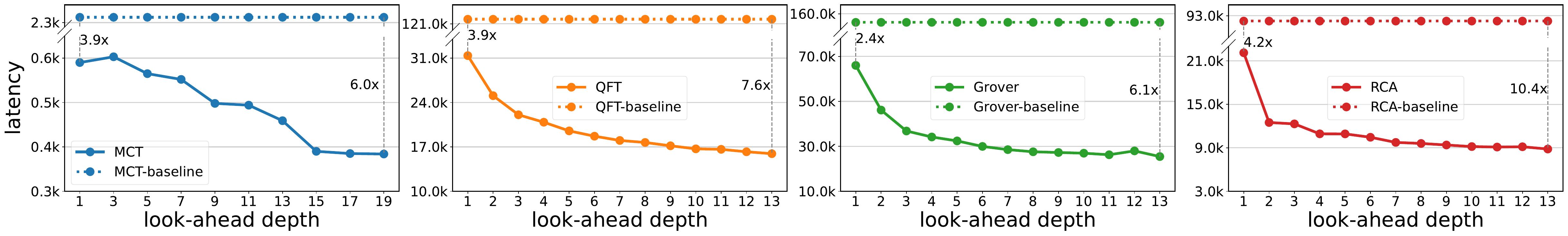}
        \\
        \vspace{-6pt}
        \hspace{20pt}{\footnotesize (b1)}\hspace{115pt}{\footnotesize (b2)}\hspace{115pt}{\footnotesize (b3)}\hspace{115pt}{\footnotesize (b4)}\\
                \vspace{-5pt}
        \caption{Performance improvement of our compiler varying with (a) buffer size and (b) look-ahead depth.}
        \label{fig:hyper_param}
\end{figure*}

\textbf{Overhead}
It can also be seen from Table~\ref{tab:main_table} that these improvements are achieved with a small overhead. On one hand, our compiler requires the generation of \todo{$7.41\%$} more (weighted) EPR pairs on average, 
which decreases as the \#QPUs per rack or \#racks increases. On the other hand, our compiler increases the wait time of generated EPR pairs in the buffer by only \todo{$6.51\times$} reconfiguration latency on average. Since this is even less than the latency of generating 1 cross-rack EPR pair, it would not lead to a significant decoherence of EPR pairs.

\begin{figure*}[bp!]
        \centering
        \includegraphics[width=1\textwidth]{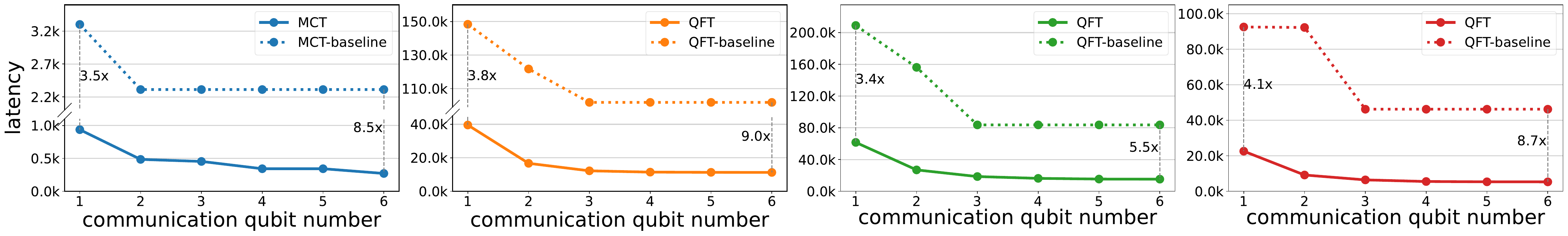}
        \\
        \vspace{-6pt}
        \hspace{20pt}{\footnotesize (a1)}\hspace{115pt}{\footnotesize (a2)}\hspace{115pt}{\footnotesize (a3)}\hspace{115pt}{\footnotesize (a4)}\\
      
        \includegraphics[width=1\textwidth]{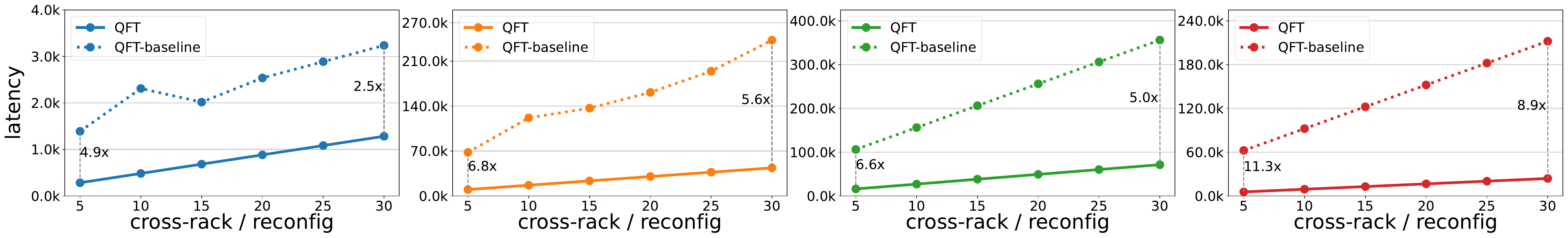}
        \\
        \vspace{-6pt}
        \hspace{20pt}{\footnotesize (b1)}\hspace{115pt}{\footnotesize (b2)}\hspace{115pt}{\footnotesize (b3)}\hspace{115pt}{\footnotesize (b4)}\\

        \includegraphics[width=1\textwidth]{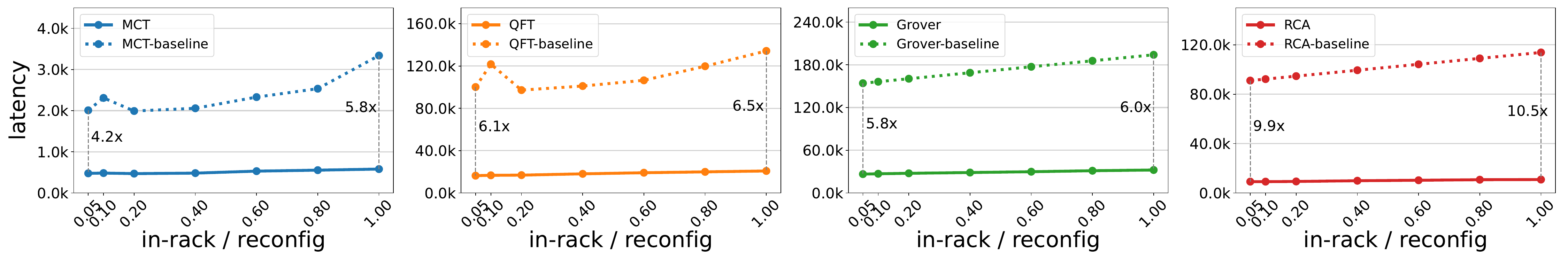}
        \\
        \vspace{-6pt}
        \hspace{20pt}{\footnotesize (c1)}\hspace{115pt}{\footnotesize (c2)}\hspace{115pt}{\footnotesize (c3)}\hspace{115pt}{\footnotesize (c4)}\\
        \caption{Performance improvement of our compiler varying with (a) \#communication qubits per QPU, (b) cross-rack EPR latency and (c) in-rack EPR latency (both normalized by reconfiguration latency).}
        \label{fig:sensitivity_analysis}
\end{figure*}
\begin{figure*}[tp!]
        \centering
        \includegraphics[width=1\textwidth]{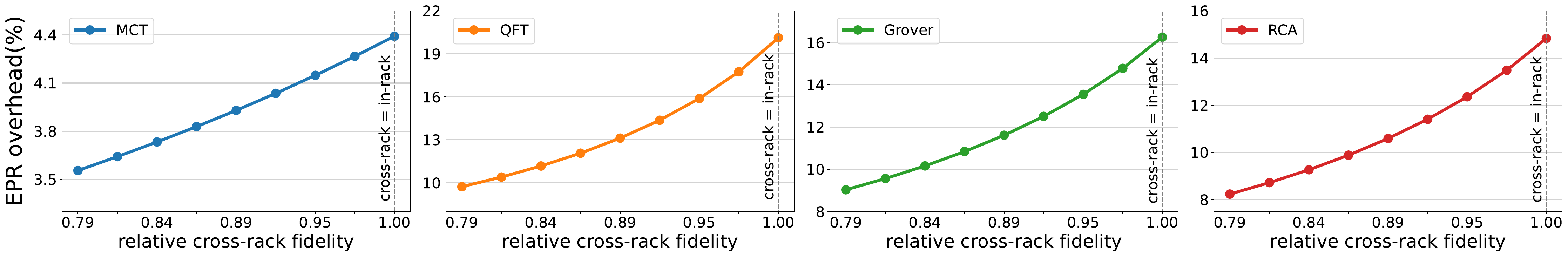}
        \\
        \vspace{-6pt}
        \hspace{20pt}{\footnotesize (a1)}\hspace{115pt}{\footnotesize (a2)}\hspace{115pt}{\footnotesize (a3)}\hspace{115pt}{\footnotesize (a4)}\\
        
        \includegraphics[width=1\textwidth]{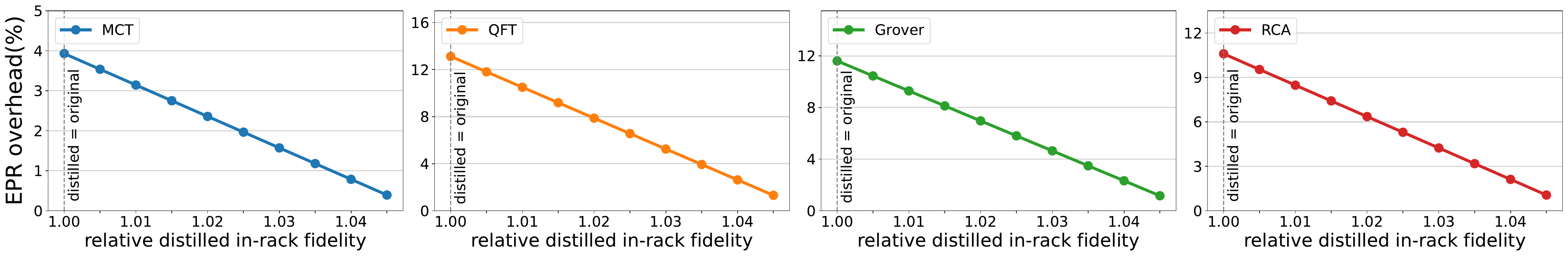}
        \\
        \vspace{-6pt}
        \hspace{20pt}{\footnotesize (b1)}\hspace{115pt}{\footnotesize (b2)}\hspace{115pt}{\footnotesize (b3)}\hspace{115pt}{\footnotesize (b4)}\\
      
        \includegraphics[width=1\textwidth]{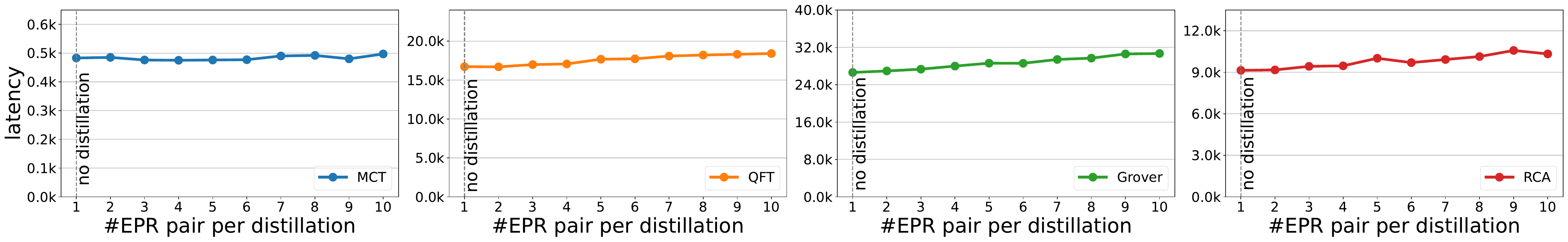}
        \\
        \vspace{-6pt}
        \hspace{20pt}{\footnotesize (c1)}\hspace{115pt}{\footnotesize (c2)}\hspace{115pt}{\footnotesize (c3)}\hspace{115pt}{\footnotesize (c4)}\\
        \caption{Fidelity overhead of our compiler varying with (a) cross-rack fidelity and (b) distilled in-rack fidelity (both relative to the original in-rack fidelity). (c) The overall latency varying with in-rack distillation through different numbers of EPR pairs.}
        \label{fig:distill}
\end{figure*}

\subsection{Choice of Hyper-parameters}
\textbf{Buffer size}
We illustrate the effect of varying buffer size while keeping all other parameters the same as program-480 in Table~\ref{tab:setting_table}. Fig.~\ref{fig:hyper_param}(a) shows the overall latency of baseline and our compiler as the buffer size increases. 
The latency of our compiler first decreases with the buffer size and then becomes stable, with the improvement factor first increasing and then becoming stable. The turning points of QFT, Grover and RCA are around \todo{7}, which takes $7/(30+7)=18.9\%$ of the total \#qubits per QPU. In contrast, the turning point of MCT is around 20, which is much larger than other benchmarks. This is because MCT is much more dominated by in-rack communications than other benchmarks. It can benefit from a larger buffer size, as in-rack EPR pairs can be collected and stored in the buffer with less restriction by network bandwidth.

\textbf{Look-ahead depth}
We also illustrate the effect of varying look-ahead depth while keeping all other parameters the same as program-480 in Table~\ref{tab:setting_table}. Fig.~\ref{fig:hyper_param}(b) shows the overall latency of baseline and our compiler as the look-ahead depth increases. The latency of our compiler first decreases with the look-ahead depth and then becomes stable, with the improvement factor first increasing and then becoming stable. Similar to buffer size, the turning point of MCT is larger than other benchmarks. It can benefit from a larger look-ahead depth, as an increased look-ahead depth leads to an increased collection of its more dominant in-rack communications.

\subsection{Sensitivity Analysis}
This subsection provides a sensitivity analysis for various hardware parameters. Since only the ratios between different latencies and fidelity matter, cross-rack and in-rack latency will be normalized by reconfiguration latency, while cross-rack and distilled in-rack fidelity will be normalized by the original in-rack fidelity.

\textbf{Communication qubit number} We illustrate the effect of varying \#communication qubits per QPU by increasing it from 1 to 6, keeping all other parameters the same as program-480 in Table~\ref{tab:setting_table}. As shown in Fig.~\ref{fig:sensitivity_analysis}(a), the overall latency of both baseline and our compiler decreases first and then becomes stable, which naturally arises from the increased bandwidth by communication qubits. The improvement factor of our compiler first increases and then becomes stable, which demonstrates its more effective utilization of network bandwidth.

\textbf{Cross-rack EPR latency}
We illustrate the effect of varying the latency of cross-rack EPR pair generation by increasing its ratio to reconfiguration latency from 5 to 30, keeping all other parameters the same as program-480 in Table~\ref{tab:setting_table}. As shown in Fig.~\ref{fig:sensitivity_analysis}(b), the overall latency of both baseline and our compiler increases with cross-rack latency.
These trends arise naturally from the fact that a longer cross-rack EPR latency leads to a longer overall latency.
The improvement factor of our compiler decreases with an increased cross-rack latency, but remains significant even if cross-rack latency is as large as $30\times$ reconfiguration latency.

\textbf{In-rack EPR latency}
We illustrate the effect of varying the latency of in-rack EPR pair generation by increasing its ratio to reconfiguration latency from 0.05 to 1, keeping all other parameters the same as program-480 in Table~\ref{tab:setting_table}. As shown in Fig.~\ref{fig:sensitivity_analysis}(c), the overall latency of both baseline and our compiler increases with in-rack latency. These trends arise naturally from the fact that a longer in-rack EPR latency leads to a longer overall latency. In contrast to cross-rack latency, the improvement factor of our compiler slightly increases with the in-rack latency.

\textbf{Relative cross-rack fidelity}
We analyze the affect of varying the fidelity of cross-rack EPR pairs, by increasing its ratio to in-rack ones from 0.79 to 1 (i.e., cross-rack fidelity from 75\% to 95\%). All other parameters are kept the same as program-480 in Table~\ref{tab:setting_table}. As shown in Fig.~\ref{fig:distill}(a), the fidelity overhead of our compiler increases as cross-rack fidelity becomes closer to in-rack fidelity. This is because our compiler accepts the tradeoff of incurring additional for hiding latencies of cross-rack communications.
With a smaller fidelity distinction between cross-rack and in-rack pairs, the cost of additional in-rack EPR pairs would appear more significant.

\textbf{Relative distilled in-rack fidelity} 
We also analyze the affect of varying the fidelity of distilled in-rack EPR pairs, as the distillation can be improved by advancing distillation protocols or sacrificing more EPR pairs for the distillation. This is demonstrated by increasing its ratio to in-rack ones from 1 to 1.047 (i.e., in-rack fidelity from 95\% to 99.5\%), keeping all other parameters the same as program-480 in Table~\ref{tab:setting_table}. As shown in Fig.~\ref{fig:distill}(b), the fidelity overhead of our compiler decreases rapidly with this fidelity ratio. That means if we can distill the additional in-rack EPR pairs to a high enough fidelity, the fidelity overhead brought by them will become negligible. 

\textbf{\#EPR pairs per distillation} If the distilled in-rack EPR fidelity is enhanced by sacrificing more EPR pairs, the generation of these EPR pairs may increase the overall latency. However, it turns out that this increase is not significant. This is because all the sacrificed EPR pairs are in-rack pairs, which can be generated collectively in our compiler. As shown in Fig.~\ref{fig:distill}(c), the overall latency is increased by only \todo{7.4\%} on average as the number of EPR pairs used in each distillation increases from 1 (i.e., no distillation) to 10.

\section{Conclusion}
In this work, we provide in-depth analysis and discussion of the compilation challenges of scaling up quantum computing with QDCs based on reconfigurable optical switch network. We identify a new optimization space to parallelize cross-rack communications by incurring minimized additional in-rack communications, proposing a flexible scheduler that decouples the generation of EPR pairs with communications while preventing deadlocks and buffer congestion caused by the flexibility.
We hope that our work could attract more effort from the computer architecture and compiler community to further explore the QDC architecture and overcome the challenges in scaling up quantum computing.

\begin{acks}

We thank the anonymous reviewers for their constructive feedback. This work is supported in part by Cisco Research, NSF 2048144, NSF 2422169, NSF 2427109. This material is based upon work supported by the U.S. Department of Energy, Office of Science, National Quantum Information Science Research Centers, Quantum Science Center. This research used resources of the Oak Ridge Leadership Computing Facility, which is a DOE Office of Science User Facility supported under Contract DE-AC05-00OR22725. The Pacific Northwest National Laboratory is operated by Battelle for the U.S. Department of Energy under Contract DE-AC05-76RL01830.
\end{acks}

\bibliographystyle{unsrturl}
\bibliography{references}


\end{document}